\documentclass[twocolumn,showpacs,preprintnumbers,amsmath,amssymb,superscriptaddress]{revtex4}
\usepackage{dcolumn}
\usepackage{graphicx,amssymb,amsmath,color}

\begin{document}

\title{Microcanonical quasi-stationarity of long-range
  interacting systems\\
 in contact with a heat bath}

\author{Fulvio Baldovin}
\affiliation{
Dipartimento di Fisica, CNISM, and
Sezione INFN, Universit\`a di Padova,
 Via Marzolo 8, I-35131 Padova, Italy
}

\author{Pierre-Henri Chavanis}
\affiliation{
Laboratoire de Physique Th\'eorique, Universit\'e Paul Sabatier,
 118 route de Narbonne, 31062 Toulouse, France
}

\author{Enzo Orlandini}
\affiliation{
Dipartimento di Fisica, CNISM, and
Sezione INFN, Universit\`a di Padova,
 Via Marzolo 8, I-35131 Padova, Italy
}

\date{\today}

\begin{abstract}
On the basis of analytical results and molecular dynamics simulations
we clarify the nonequilibrium dynamics of a long-range interacting
system in contact with a heat bath.  For small couplings with the
bath, we show that the system can first be trapped in a Vlasov
quasi-stationary state, then a microcanonical one follows, and finally
canonical equilibrium is reached at the bath temperature. We
demonstrate that, even out-of-equilibrium,
Hamiltonian reservoirs microscopically coupled with the system and
Langevin thermostats provide equivalent descriptions. Our
identification of the key parameters determining the quasi-stationary
lifetimes could be exploited to control experimental systems such as
the Free Electron Laser, in the presence of external noise or inherent
imperfections.
\end{abstract}

\maketitle

In recent years, systems characterized by interactions that slowly decay
at large distances have considerably attracted the attention of
experimental and theoretical physicists. Plasmas, gravitational
systems, two-dimensional vortices, wave-matter
interaction systems, all fall in this category \cite{houches}.  
Of particular
interest to what follows is the case of the Free Electron Laser (FEL),
a source of coherent radiation which is expected to outperform
traditional lasers thanks to the properties of a relativistic-electron
lasing medium (see, e.g., \cite{fel} and references therein).
For these systems, the prevalence of long-range interactions over
mechanisms acting on short-range scales implies an inefficiency of fast
collision processes.  This is in contrast with the assumptions
underlying Boltzmann's derivation of the transport equation and brings
to the fact that long-range interacting systems get easily stuck in
(Vlasov) nonequilibrium quasi-stationary states (QSS)
\cite{houches,mukamel,epjb,anto,anto_1}. For instance, the possible observation of QSSs in FEL
experiments is predicted on the basis of molecular dynamics
simulations of isolated (microcanonical) Hamiltonian systems capturing
the essential features of the FEL dynamics \cite{barre}. From an
experimental point of view, it is crucial to recognize if a stable
nonequilibrium picture survives in the presence of an external
environment (or inherent imperfections) acting on the system
\cite{baldovin,bbgky,choi}, and which are the parameters playing a key role in
the determination of the QSSs lifetimes.  
In addition, a basic theoretical issue
is whether a stochastic dynamics simulating a thermal
bath (TB), e.g., of the Langevin type, reproduces the same
nonequilibrium features of a Hamiltonian reservoir microscopically
interacting with the long-range system.
 
While for short-range systems
the equivalence between Hamiltonian and
Langevin thermostats is well established, the connection between these
two different descriptions for the nonequilibrium
behavior of a long-range system is less clear, and recent
simulations \cite{baldovin_1} recovered the same results for the 
two TBs only at equilibrium.
Here we demonstrate that
 Hamiltonian and Langevin TB provide in fact an equivalent
description of the behavior of a paradigmatic
long-range system also in nonequilibrium conditions. 
This is established, both analytically and
numerically, analyzing the scaling properties of the QSSs
lifetimes. 
We recast the interaction between the system and a 
Hamiltonian TB in terms of an equivalent set of
generalized Langevin equations. The damping coefficient $\gamma$
determines the timescale $t_{bath}$ for the relaxation to canonical equilibrium, achieved when system and 
TB share the same temperature. 
However, even in
the presence of  TB, correlations due to 
a slow collisional process determine 
another timescale $t_{coll}$ (depending on system size $N$)
which corresponds to a relaxation  
to a {\it  microcanonical QSS}. 
Thus, for $\gamma$ small enough and $N$
not too large, our main result is the discovery of 
a novel picture for the transport to equilibrium: On a timescale $t_{d y
n}\sim1$ (in dimensionless
units) a violent relaxation drives the system
into a Vlasov QSS; Then, on a timescale $t_{coll}\sim N^\delta$
(with $\delta\ge 1$) 
the system reaches a microcanonical QSS; Finally,
on a timescale $t_{bath}\sim1/\gamma$, the system crosses over to
canonical (thermal) equilibrium.

Thanks to its appealing (yet non-trivial) simplicity, a system which
captured a paradigmatic interest among the researchers is the
Hamiltonian Mean Field (HMF) model
\cite{houches,epjb,anto,anto_1,baldovin,bbgky,choi}, which can be thought as a set 
of $N$  globally coupled $X Y$-spins
with Hamiltonian
\begin{equation}
H_{HMF}=\sum_{i=1}^N\frac{l_i^2}{2}
+\frac{1}{2N}\sum_{i,j=1}^N\left[1-\cos(\theta_i-\theta_j)\right],
\label{hmf}
\end{equation}
where $\theta_i\in[0,2\pi)$ are the spin angles and
$l_i=\dot\theta_i\in\mathbb R$ their angular momenta
(velocities). Defining the kinetic temperature
$T\equiv \sum_{i=1}^N l_{i}^{2}/N$ as twice the specific kinetic energy and the
specific magnetization as $m\equiv|\sum_{i=1}^N(\cos\theta_i,\sin\theta_i)|/N$,
one obtains the exact relation $E/N=T/2+(1-m^2)/2$,
where $E$ is the total energy of the system.  The free energy of the
HMF model can be exactly mapped \cite{barre} onto that of the Colson-Bonifacio
Hamiltonian model for the single-pass 
FEL \cite{colson}. 
In such a context, the variables $l_i$'s are
interpreted as the phase velocities relative to the center of mass of
the $N$ electrons and the $\theta_i$'s are
the electron phases with respect to
the co-propagating wave \cite{barre}. 
Despite some dynamical and thermodynamic differencies,  
analogies can also be found between the HMF model and 
the behavior of one-dimensional self-gravitating systems \cite{miller}.

When the HMF model is
isolated, in order to determine whether the system truly converges
toward statistical equilibrium and the timescale of this relaxation,
one must develop an appropriate kinetic theory.
This is a classical problem
addressed, e.g., in \cite{newref} and, more recently, in  
\cite{bbgky}. 
From the Liouville
equation it is possible 
to derive the 
BBGKY hierarchy for the reduced joint probability density functions
(PDF)
$p_{j}(\theta_1,\ldots,\theta_{j},\dot\theta_1,\ldots,\dot\theta_{j},t)$,
with ${j}=1,2,\ldots,N$.  In the thermodynamic limit
$N\to\infty$, $E/N\sim1$, and fixed volume $V=2\pi$, the hierarchy can
be closed by considering a systematic expansion in
powers of $1/N$ of the solutions of the equations
\cite{bbgky}. 
At the order $1/N$, the 
distribution function $f=Np_{1}$ 
satisfies 
\begin{equation}
\frac{\partial f}{\partial t}
+\dot\theta\frac{\partial f}{\partial\theta}
-\frac{\partial\phi}{\partial\theta}
\frac{\partial f}{\partial\dot\theta}
=\frac{1}{N}C_{coll}(f),
\label{eq_fmicro}
\end{equation} 
where 
$\phi(\theta,t)\equiv-\frac{1}{2\pi}
\int_0^{2\pi}d\theta^\prime
\cos(\theta-\theta^\prime)
\int_{-\infty}^{+\infty}d\dot\theta\
f(\theta^\prime,\dot\theta,t)$.  For $N\rightarrow +\infty$, the
r.h.s. of Eq. (\ref{eq_fmicro}) is negligible and we get the Vlasov
equation \cite{newref}. For $N$ finite, $C_{coll}(f)$ is a
``collision'' operator taking into account
correlations between particles \cite{bbgky}. The scaling in
Eq. (\ref{eq_fmicro}) indicates that ``collisions'' (more properly
correlations) operate on a very slow timescale, of the order $N t_{d y
n}$ or even larger when $C_{coll}(f)= 0$, as for a spatially
homogeneous one dimensional system \cite{bbgky}. It is precisely
because of the development of correlations that the
system reaches, on the ``collisional'' relaxation time $t_{coll}$, a
microcanonical Boltzmann distribution.
Since $t_{coll}(N)$ diverges for $N\rightarrow+\infty$ 
(different scalings, 
$t_{coll}\sim N t_{d y n},N^{1.7}t_{d y n},e^{N}t_{d y n}$, have been reported
depending on the initial conditions \cite{timescale}), 
the domain of validity of the Vlasov equation is huge.
Starting from an out-of-equilibrium initial condition, the Vlasov
equation develops a complicated mixing process in the single-particle
phase space, leading, in most cases, to a QSS \cite{anto_1,timescale}. 
This process is called
violent relaxation since it takes place on a timescale $t_{d y
n}\sim1$. The QSS is a nonlinearly dynamically stable stationary
solution of the Vlasov equation on a coarse-grained scale
\cite{bbgky}.
The Vlasov equation admits an infinite number of stationary solutions.
The statistical theory of Lynden-Bell \cite{lb} predicts the ``most
probable'' (most mixed) state \cite{epjb,anto}. However, in view of
the possible occurrence of incomplete relaxations \cite{bbgky,epjb},
there are cases in which the QSSs take forms different from those
described by Lynden-Bell's theory.

\begin{figure}
\includegraphics[width=0.70\columnwidth]{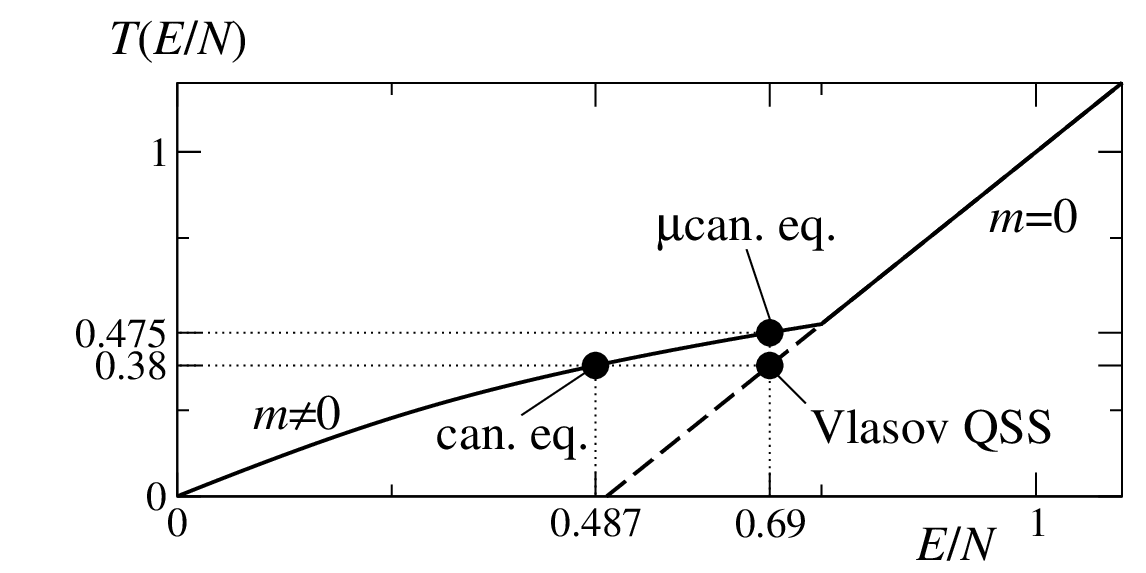}
\caption{
  Caloric curve of the HMF model (solid line). The 
  dashed line is the prolongation of the ordered
  phase to subcritical energies.
  See text for details.
}
\label{fig_caloric}
\end{figure}

The interaction of the system (\ref{hmf})
with a reservoir has been studied 
in Refs. \cite{baldovin} and \cite{bbgky}
introducing a Hamiltonian and a Langevin  
TB, respectively. 
In the latter case, the dynamics of the system is governed by a set of
$N$ coupled stochastic equations:
\begin{equation}
\ddot\theta_i=F_i
-\gamma\dot\theta_i+\sqrt{2\gamma {T}_{b}}\xi_i(t),
\label{eq_lange}
\end{equation} 
where $F_i\equiv-\frac{1}{N}\sum_{j=1}^N\sin(\theta_i-\theta_j)$ is
the long-range force experienced by the spin $i$, $\gamma$ is a
damping coefficient due to the interaction with the TB, ${T}_{b}$ is
the TB temperature, and $\xi_i$ is a Gaussian white noise satisfying
$\langle\xi_i(t)\rangle=0$ and
$\langle\xi_i(t)\xi_j(t^\prime)\rangle=\delta_{i j}\delta(t-t^\prime)$.
Eqs. (\ref{eq_lange}) define the so-called Brownian Mean Field model
\cite{bbgky} (see also \cite{choi}). 
The evolution of
the $N$-body PDF is governed in this case  
by the Fokker-Planck equation, 
from which a BBGKY-like hierarchy for the $p_{j}$'s can
be derived.  In the thermodynamic limit $N\rightarrow
+\infty$, $T_{b}\sim 1$, $V=2\pi \sim 1$, the hierarchy can be closed
by considering an expansion in powers of $1/N$.
It is possible 
to see \cite{bbgky} that 
$p_2(\theta_1,\dot\theta_1,\theta_2,\dot\theta_2)=
p_1(\theta_1,\dot\theta_1)\;p_1(\theta_2,\dot\theta_2)+O(1/N)$. 
Hence, for $N\rightarrow +\infty$, the equation for $f=Np_{1}$ becomes
the mean-field Kramers equation:
\begin{equation}
\frac{\partial f}{\partial t}
+\dot\theta\frac{\partial f}{\partial\theta}
-\frac{\partial\phi}{\partial\theta}
\frac{\partial f}{\partial\dot\theta}
=\gamma
\frac{\partial}{\partial \dot\theta}\left ({T_{b}}\frac{\partial f}{\partial  \dot\theta}+f\dot\theta\right )\equiv \gamma C_{bath}(f).
\label{eq_single}
\end{equation}
This equation
relaxes to the canonical mean-field
Maxwell-Boltzmann distribution
$f(\theta,\dot\theta)=\frac{1}{Z}e^{-(\dot\theta^2/2+\phi(\theta))/T_b}$
[with $\phi(\theta)=\lim_{t\to\infty}\phi(\theta,t)$] on a timescale
$t_{bath}\sim1/\gamma$, independent of $N$.  For $\gamma=0$
one recovers the microcanonical situation in which the system is
isolated. Correspondingly, the Fokker-Planck equation  becomes the Liouville
equation and the mean field Kramers equation becomes the Vlasov equation.

\begin{figure}
\includegraphics[width=0.76\columnwidth]{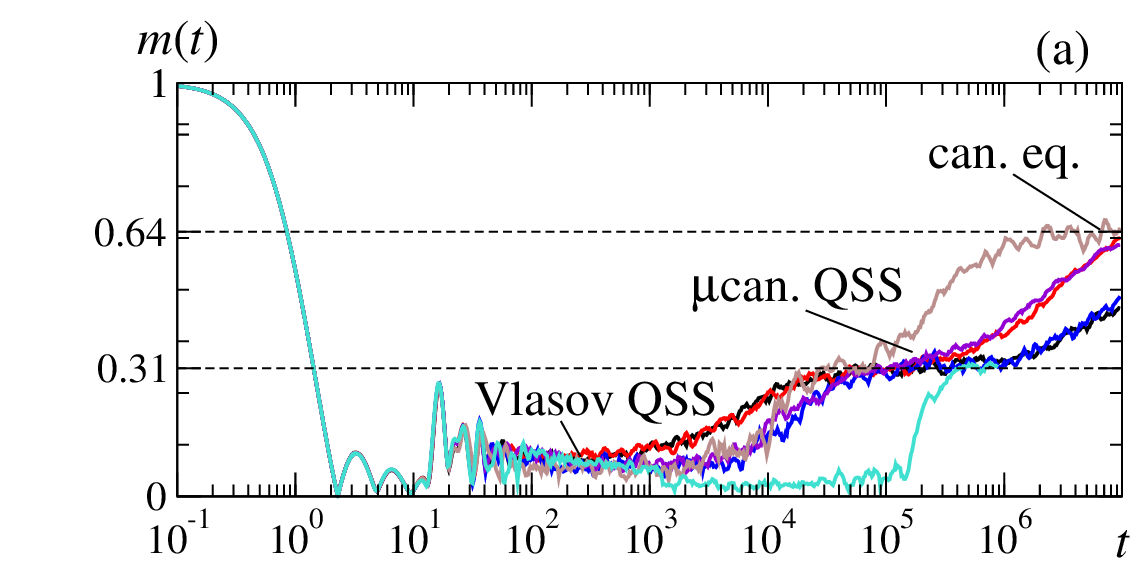}\\
\vspace{0.25cm}
\includegraphics[width=0.33\columnwidth]{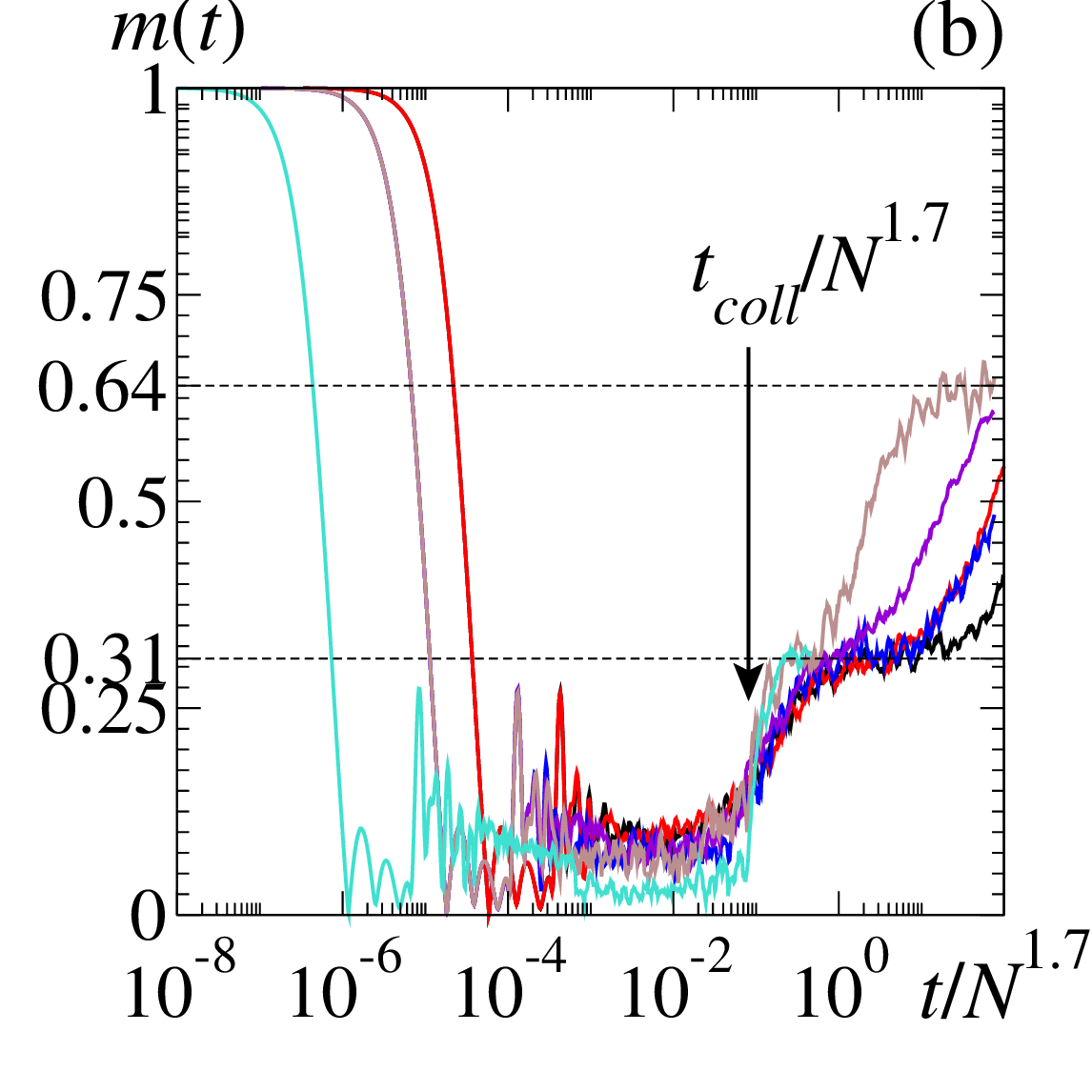}
\includegraphics[width=0.33\columnwidth]{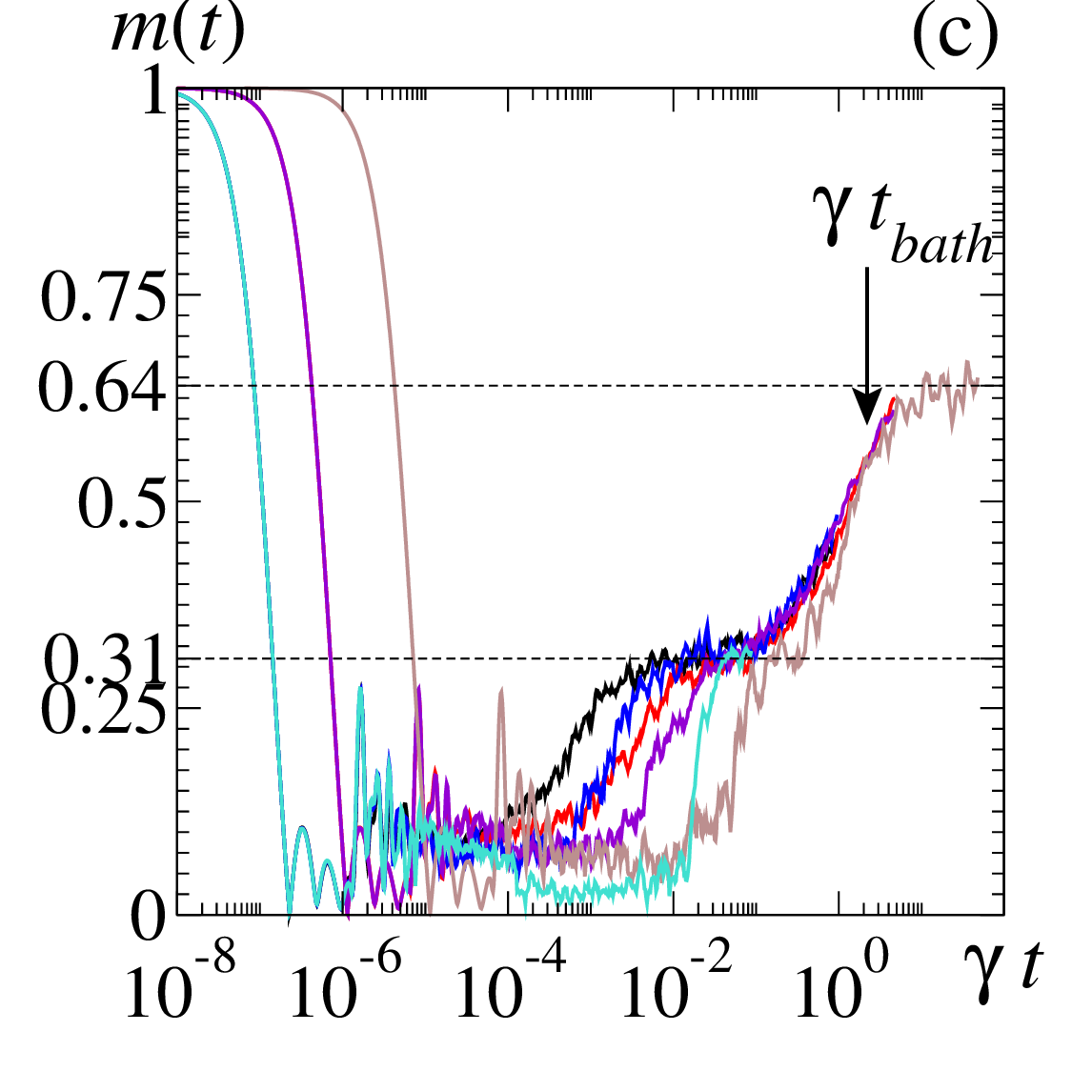}\\
\caption{ (Color online) Time evolution of the magnetization with 
  a Langevin TB. The control parameters 
  $\gamma=10^{-7},5\times10^{-7},10^{-6}$, $N=500,1000,5000$ satisfy
  $\gamma\ll 1/N^{1.7}\ll 1$. Plots are averages over at most $10$
  runs.
  The system reaches a Vlasov QSS for $t\sim t_{d y n}\sim 1$, a
  microcanonical QSS for $t\sim t_{coll}\sim N^{1.7}$ and a
  canonical equilibrium for $t\sim t_{bath}\sim 1/\gamma$. 
}
\label{fig_lange}
\end{figure}

The numerical integration of Eqs. (\ref{eq_lange}) exhibits a very
rich transport-to-equilibrium picture \cite{baldovin_1}. In this Letter, we show that
the key point for understanding the nonequilibrium behavior of the
system is the comparison between the timescales $t_{d y n}$, $t_{coll}$
and $t_{bath}$. We specifically analyze the time evolution of the
magnetization of the system for simulations with random water bag initial
conditions of the form: $p_1(\theta,\dot\theta,0)=\delta(\theta-0)
[\vartheta(\dot\theta+\overline l)+\vartheta(\dot\theta-\overline l)]
/2\overline l$ ($\vartheta$ being the Heaviside step function), where
$\overline l\simeq2.03$. We thus have $m(0)=1$, $E(0)/N\simeq0.69$,
and $T(0)=1.38$. These and similar nonequilibrium initial conditions have been
largely studied in microcanonical simulations \cite{anto_1,timescale,anto} and
recently discussed in the presence of a TB \cite{baldovin,baldovin_1}.  
The initial energy of the system is below the critical point $E_{c}/N=3/4$
(see Fig. \ref{fig_caloric}). Microcanonical simulations ($\gamma=0$)
display, in a time $t_{d y n}\sim1$, a violent relaxation process in
which the magnetization drops to $m\simeq0+O(1/\sqrt{N})$ and the temperature
to $T\simeq0.38$. A QSS lasting a time of the order $t_{coll}\sim
N^{\delta}$ 
follows the violent relaxation
\footnote{
For this $m(0)=1$ initial condition, the Vlasov QSS is {\it not}
described by the Lynden-Bell theory because of incomplete
relaxation (see \cite{bbgky,epjb}).
}. After the QSS,
the isolated system warms up (at fixed energy) due to finite-size
effects, and finally reaches the microcanonical equilibrium state with
$T\simeq 0.475$ and $m\simeq0.31$. At variance, in a relaxation
process at fixed temperature $T_b=0.38$ (canonical simulations with $\gamma\neq0$), the
system reaches a canonical equilibrium state with $T=T_b$,
$E/N\simeq0.487$, and $m\simeq0.64$ 
(see again Fig. \ref{fig_caloric}). 
Hence, if we fix the TB temperature in
Eq. (\ref{eq_lange}) at $T_b=0.38$, and let $\gamma\to0$,
there is an apparent discontinuity in the final equilibrium value of
the magnetization \cite{baldovin_1}. Actually, this paradox is solved
by the presence of a second QSS which follows the Vlasov one. 
Indeed, for $t\ll t_{bath}\sim 1/\gamma$ the energy is relatively well
conserved. Thus, 
if $t_{d y n}\ll t_{coll}\ll t_{bath}$
(i.e. $\gamma\ll1/N^\delta\ll 1$), the magnetization of the system
relaxes to the microcanonical value $m\simeq 0.31$ on the collisional
timescale $t_{coll}\sim N^\delta$ (we find $\delta\simeq1.7$,
independently of $\gamma$).  
This is the reason why we call this second
quasi-equilibrium state the ``microcanonical QSS''.
The equilibrium with the TB, and the
consequent value $m\simeq0.64$, is established only on the much larger
timescale $t_{bath}$.  
On the contrary, for $t_{d y n}\ll t_{bath}\ll t_{coll}$
(i.e. $1/N^{\delta}\ll\gamma\ll 1$), the system first reaches a Vlasov
QSS on a timescale $t_{d y n}$, then a canonical equilibrium state with
temperature $T=T_{b}$ on a timescale $t_{bath}$, and does not form a
microcanonical QSS. We note that in order to see the microcanonical
QSS we need a very small noise level: $\gamma\ll 1/N^{\delta}$.
There is a further interesting situation, obtained for $t_{bath}\ll
t_{d y n}$ (i.e. $\gamma\gg 1$). In this latter case, the system reaches
a canonical equilibrium state with temperature $T=T_{b}$, without
forming any (Vlasov or microcanonical) QSS. This corresponds to the
overdamped (Smoluchowski) regime studied in \cite{bbgky}. In
conclusion, the limits $t\to\infty$, $N\to\infty$, $\gamma\to 0$ do
not commute. Depending on the order in which they are taken, the
average value of the magnetization can be the Vlasov 
($N\rightarrow \infty$ and $\gamma\rightarrow 0$ before $t\rightarrow \infty$), 
the microcanonical 
($\gamma\rightarrow 0$ and $t\rightarrow \infty$ before
$N\rightarrow \infty$), or the canonical one  
($N\rightarrow \infty$ and $t\rightarrow \infty$ before 
$\gamma\rightarrow 0$). The simulations reported in
Fig. \ref{fig_lange} demonstrate these features.
In Fig. \ref{fig_lange}a curves with the same $N$ almost coincide 
for $t<t_{coll}$; Those with the same $\gamma$ collapse onto each
other for $t>t_{coll}$. 
The presence of a microcanonical QSS
following the Vlasov one is particularly evident in the rescaled plots
of Figs. \ref{fig_lange}b,c, which confirm the scaling
properties of $t_{coll}$ and $t_{bath}$.

The next step is to establish whether these microcanonical QSSs are an artifact
of the mesoscopic stochastic dynamics (\ref{eq_lange}) 
or if they are still present when
we consider a Hamiltonian TB microscopically coupled with the long-range system. 
In Ref. \cite{baldovin} 
a first-neighbors coupled $X Y$-spins TB has been introduced,
\begin{equation}
H_{TB}=\sum_{i=1}^{N_{TB}}\frac{{l_i^{TB}}^2}{2}
+\sum_{i=1}^{N_{TB}}\left[1-\cos(\theta_{i+1}^{TB}-\theta_i^{TB})\right],
\label{eq_tb1}
\end{equation}
interacting with the HMF model through the potential
\begin{equation}
H_{I}=\epsilon\sum_{i=1}^{N}\sum_{s=1}^S\left[1-\cos(\theta_{i}-\theta_{r_s(i)}^{TB})\right], 
\label{eq_interaction1}
\end{equation}
where $r_s(i)$ are integer independent random numbers in the interval
$[1,N_{TB}]$. Each HMF-spin is in contact with a set of
$S$ different TB-spins chosen randomly along the chain, and the
coupling constant $\epsilon\geq0$ determines the interaction strength
between system and TB. The conditions \cite{baldovin} $N_{TB}=N^2$ and
$S=10^5 N^{-1/2}$ assure that for large $N$ the
interaction, the system, and the TB energies are well
separated.  
Molecular dynamics
simulations of the Hamiltonian $H_{HMF}+H_{TB}+H_I$ 
were shown to agree with the Langevin ones at equilibrium
\cite{baldovin_1}, 
whereas the
presence of QSSs with a lifetime depending on both $\epsilon$ and $N$
(or, equivalently, $S$)   
has been detected \cite{baldovin,baldovin_1}.  

In order to
clarify this dependence we study a different
Hamiltonian form for the TB and for its interaction with the system,
which has the advantage of allowing an explicit analytical
analysis. Following the approach outlined by Zwanzig \cite{zwanzig}
for short-range systems, our aim
is to recast the Hamiltonian equations of motion 
in a form similar to Eqs. (\ref{eq_lange}). Hence, we replace
Eqs. (\ref{eq_tb1},\ref{eq_interaction1}) with
\begin{equation}
H_{TB}^\prime=\sum_{i=1}^{N_{TB}}\frac{{l_i^{TB}}^2}{2},\;
H_{I}^\prime=\epsilon\sum_{i=1}^{N}\sum_{s=1}^S\left[
\frac{\omega_{r_s(i)}}{4}\left(\theta_{i}-\theta_{r_s(i)}^{TB}\right)^2\right], 
\label{eq_interaction2}
\end{equation}
respectively, and consider the Hamiltonian
$H_{HMF}+H_{TB}^\prime+H_I^\prime$. 
The TB in
Eqs. (\ref{eq_interaction2})
describes a set of $N_{TB}$ isochronous harmonic oscillators in their
canonical coordinates, which interact with the system through a
quadratic potential. 
This quadratic form
can be thought as a small-angles expansion of
Eq. (\ref{eq_interaction1}), and again each element of the HMF model
interacts with $S$ different TB oscillators. 
Using for example the Laplace transform and performing then
an integration by parts, 
the Hamiltonian dynamics of the $\theta_i$'s becomes 
\begin{equation}
\ddot\theta_i(t)=F_i
-\int_0^t d t K(t-\tau)\;\dot\theta_i(\tau)
+\xi^\prime(t),
\label{eq_zwa}
\end{equation}
where 
$K(t)\equiv\epsilon\sum_{i=1}^S\omega_{r_s(i)}^2\cos\left[
\sqrt{\epsilon}\;\omega_{r_s(i)}t\right]$, and 
$\xi^\prime$ can be written explicitly in terms of the initial
conditions 
\footnote{ 
$\xi^\prime(t)\equiv\epsilon\sum_{i=1}^S\omega_{r_s(i)}^2\left\{
\left[\theta_{r_s(i)}^{TB}(0)-\theta_i(0)\right]
f_1(t)
+\dot\theta_{r_s(i)}^{TB}(0)
f_2(t)
\right\}$,
where $f_1(t)\equiv
\cos\left(\sqrt{\epsilon}\;\omega_{r_s(i)}t\right)
$,
and
$f_2(t)\equiv
\sin\left(\sqrt{\epsilon}\;\omega_{r_s(i)}t\right)/\sqrt{\epsilon}\;\omega_{r_s(i)}
$.
}.
\begin{figure}
\vspace{0.4cm}
\includegraphics[width=0.33\columnwidth]{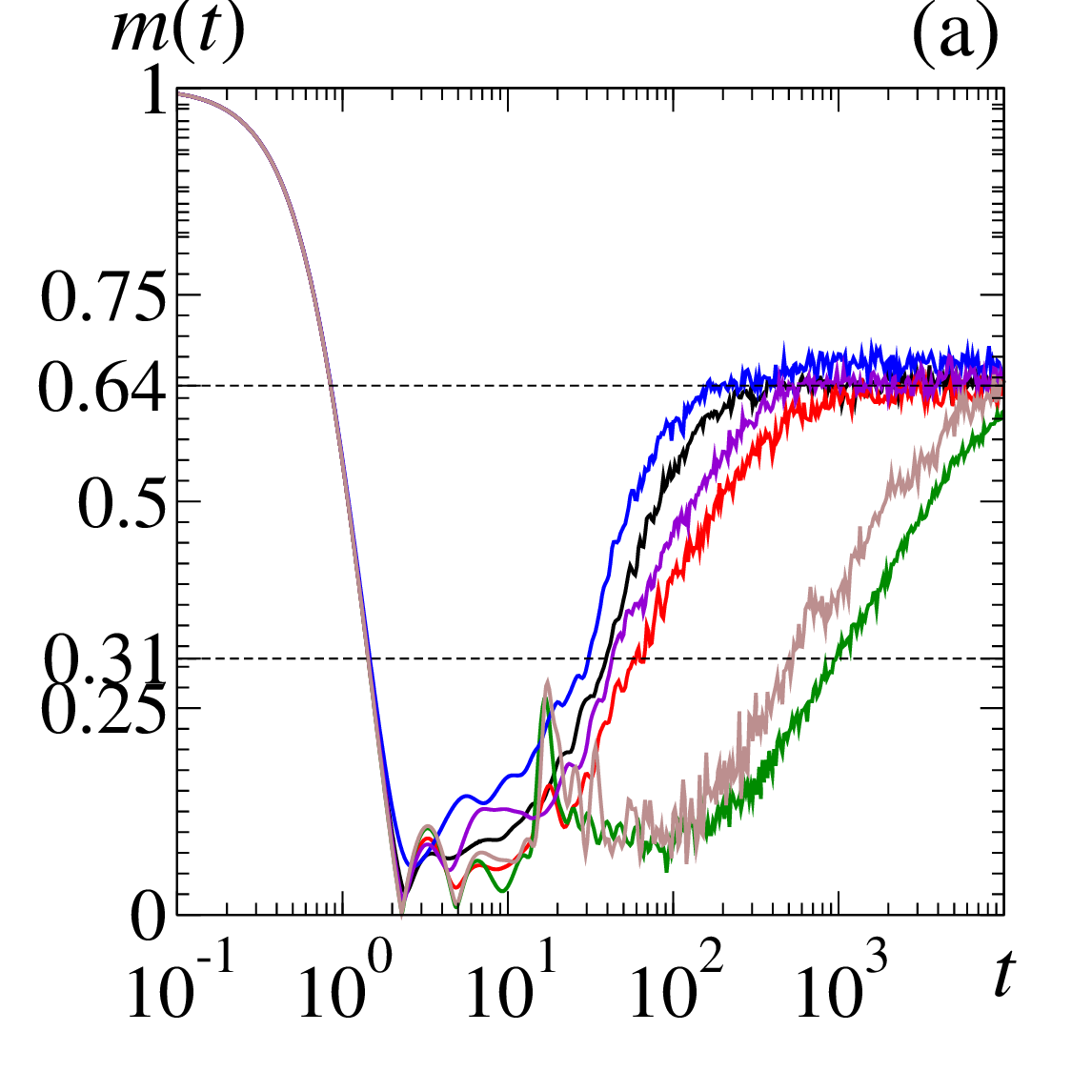}
\includegraphics[width=0.33\columnwidth]{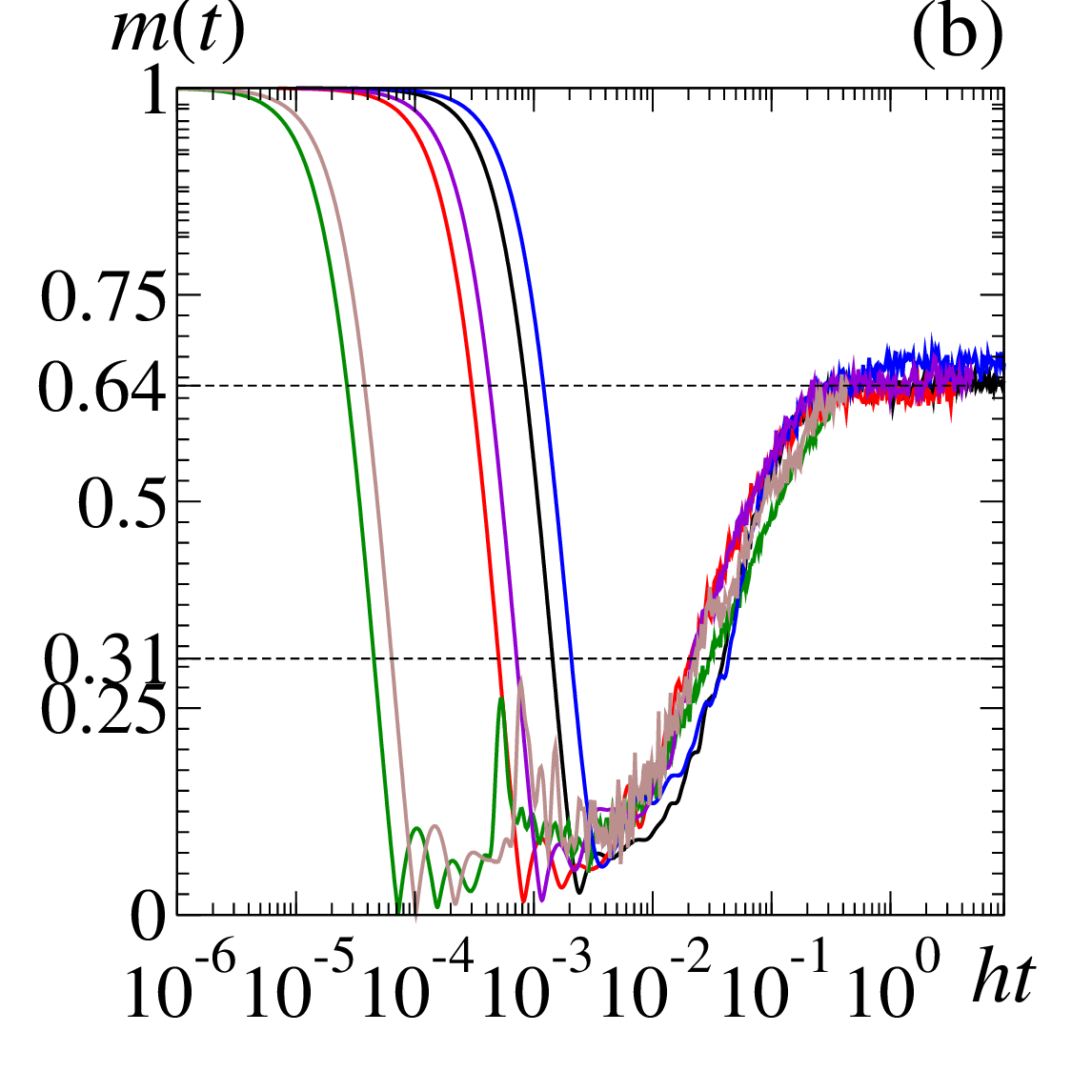}\\
\includegraphics[width=0.76\columnwidth]{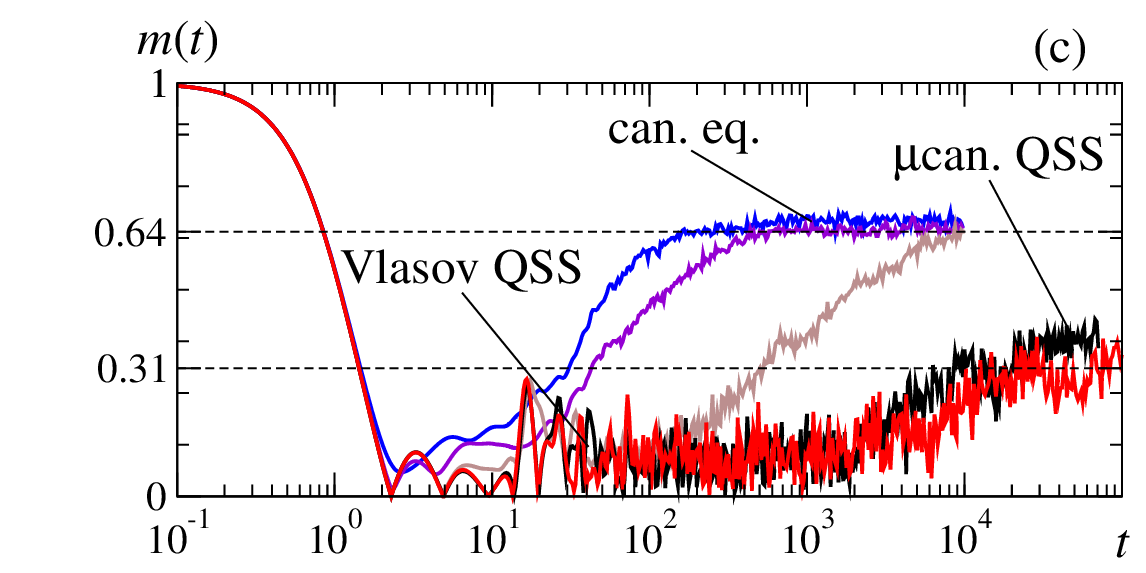}\\
\caption{
  (Color online) Time evolution of $m$ with the Hamiltonian TB in
  Eqs. (\ref{eq_tb1},\ref{eq_interaction1}).  In (a,b) the control
  parameters are $\epsilon=0.1,0.05,0,01$; $N=500,1000,5000$;
  $S=10^5N^{-1/2}$. We found $h\equiv\epsilon^{3/2}S$.  In (c) the
  longer simulations are with $\epsilon=10^{-3},10^{-4}$;
  $N=500$. Averages are over at most $10$ runs.  }
\label{fig_ham}
\end{figure}
Assuming a random distribution for the initial data, 
$\xi^\prime$ can be regarded as a stochastic term.
On the other hand, $K$ is a memory kernel which depends on $\epsilon$,
$S$, and on the distribution of the frequencies $\omega_{r_s(i)}$.
Specifically, when $K$ reduces to a $\delta$-function, 
Eq. (\ref{eq_zwa}) recasts into Eq. (\ref{eq_lange}) with 
$\gamma=h(\epsilon,S)$, and $h$ a model-dependent function.
The form of equations (\ref{eq_zwa}) 
suggests that
the relaxation process in the presence of a general Hamiltonian TB  
should be analogous to that described by the stochastic 
Langevin Eqs. (\ref{eq_lange}), with the
canonical equilibrium established on a timescale
$t_{bath}=[h(\epsilon,S)]^{-1}$.
In particular, by choosing
sufficiently small $\epsilon$'s, 
the system should exhibit microcanonical QSS
also if coupled with a Hamiltonian TB. 
We verified these conclusions for the Hamiltonian TB in  
Eqs. (\ref{eq_tb1},\ref{eq_interaction1}).
Figs. \ref{fig_ham}a,b demonstrate that if we rescale the time by 
$h(\epsilon,S)=\epsilon^{3/2}S$ indeed the relaxation time to 
the thermal equilibrium obtained for different 
$\epsilon$'s and $S$'s collapse onto the same value. 
Moreover, for $\epsilon\leq10^{-3}$ 
microcanonical QSSs clearly appear (Fig. \ref{fig_ham}c).

In summary, we have shown that the nonequilibrium dynamics of a
paradigmatic long-range system which can be mapped onto the one
describing the single-pass FEL is characterized by the three
timescales $t_{d y n}\sim1$, $t_{coll}\sim N^\delta$, $t_{bath}\sim
1/\gamma$. By acting on the initial conditions, on the system size $N$,
or on the coupling with the heat bath $\gamma$, one can conceive
experiments in which the system is either in equilibrium with the bath
or in a quasi-equilibrium state with a dynamical temperature which is 
different from the temperature of the thermal environment. 
This situation
could inspire interesting applications and provides a control on the
imperfections influencing a FEL and other long-range systems.


\begin{thebibliography}{99}

\bibitem{houches} 
{\it Dynamics and thermodynamics of systems with long range
interactions}, eds T. Dauxois {\it et al.}, Lect. Notes Phys. 
{\bf 602} (Springer, 2002); 
{\it Dynamics and Thermodynamics of Systems with Long-range
  Interactions: Theory and Experiments}, eds A. Campa {\it et al.}, 
AIP Conf. Proc. {\bf 970} (AIP, 2008).

\bibitem{fel}
E. Pl\"onjes {\it et al.}, Phys. World {\bf 16}(7), 33 (2003);
S. Milton {\it et al.}, Science {\bf 292}, 2037 (2001);
E. Allaria and G. De Ninno, Phys. Rev. Lett. {\bf 99}, 018801 (2007).

\bibitem{mukamel}
D. Mukamel {\it et al.}, Phys. Rev. Lett. {\bf 95}, 240604 (2005).

\bibitem{epjb} 
P.H. Chavanis, Eur. Phys. J. B {\bf 53}, 487 (2006).

\bibitem{anto} 
A. Antoniazzi {\it et al.} Phys. Rev. E
{\bf 75}, 011112 (2007).

\bibitem{anto_1}
A. Antoniazzi {\it et al.}, Phys. Rev. Lett. {\bf 98}, 150602 (2007);
A. Antoniazzi {\it et al.}, Phys. Rev. Lett. {\bf 99}, 040601 (2007).

\bibitem{barre}
J. Barr\'e {\it et al.}, Phys. Rev. E {\bf 69}, 045501(R) (2004);
J. Barr\'e {\it et al.}, J. Stat. Phys. {\bf 119}, 677 (2005).

\bibitem{baldovin}
F. Baldovin and E. Orlandini, Phys. Rev. Lett. 
{\bf 96}, 240602 (2006); {\bf 97}, 100601 (2006).

\bibitem{bbgky}  
P.H. Chavanis {\it et al.}, Eur. Phys. J. B {\bf 46}, 61 (2005);
P.H. Chavanis, Physica A {\bf 361}, 81 (2006); {\bf 387}, 787 (2008);
P.H. Chavanis and C. Sire {\bf 73}, 066104 (2006).

\bibitem{choi}  
M.Y. Choi, J. Choi, Phys. Rev. Lett. {\bf 91}, 124101 (2003).

\bibitem{baldovin_1}
F. Baldovin and E. Orlandini, Int. J. Mod. Phys. B {\bf 21}, 4000 (2007).

\bibitem{colson}
W. B. Colson, Phys. Lett. A {\bf 59}, 187 (1976);
R. Bonifacio {\it et al.}, Opt. Commun. {\bf 50}, 373 (1984).

\bibitem{miller}
K.R. Yawn, B. Miller, Phys. Rev. Lett. 79, 3561 (1997).

\bibitem{newref}  
W. Braun, K. Hepp, Comm. Math. Phys. {\bf 56}, 101 (1977);
J. Messer, H. Spohn, J. Stat. Phys. {\bf 29}, 561 (1982).

\bibitem{timescale}  
V. Latora {\it et al.}, Phys. Rev. E {\bf 64}, 056134 (2001);
Y.Y. Yamaguchi {\it et al.}  Physica A {\bf 337}, 36 (2004); A. Campa
{\it et al.} Phys. Rev. E {\bf 76}, 041117 (2007);
K. Jain {\it et al.}, J. Stat. Mech., P11008 (2007). 


\bibitem{lb}  
D. Lynden-Bell, MNRAS  {\bf 136}, 101 (1967).

\bibitem{zwanzig}
R. Zwanzig, {\it Nonequilibrium Statistical Mechanics} 
(Oxford University Press, 2001).




\end{thebibliography}
\end{document}